\documentclass[10pt]{article}

\usepackage{graphicx}
\usepackage{fullpage}
\usepackage{latexsym}
\usepackage{eepic}
\usepackage{times}
\usepackage{natbib}

\begin{document}

\def\degC{$^\circ$C}
\def\phiphage{$\phi29$}
\def\lamphage{$\lambda$}
\def\lamphageC{$\lambda$cI60}
\def\lamphageB{$\lambda$b221}
\def\lamphageBlong{$\lambda$b221cI26}
\def\ecoli{\textit{E.~coli}}
\def\shigella{\textit{S.~sonnei}}
\def\Rin{R_{\rm in}}
\def\Rout{R_{\rm out}}
\def\ds{d_{\rm s}}

\def\nm{\textrm{nm}}
\def\bp{\textrm{bp}}
\def\atm{\textrm{atm}}
\def\pN{\textrm{pN}}
\def\nm{\textrm{nm}}
\def\ul{\textrm{$\mu$L}}
\def\ug{\textrm{$\mu$g}}
\def\kB{k_{\rm B}}
\def\RDNA{R_{\rm DNA}}

\def\mPEG{m_{\rm PEG}}
\def\mtot{m_{\rm tot}}

\title{The effect of genome length on ejection forces\\ in bacteriophage lambda}
\author{Paul Grayson$^{1,\dagger}$,
Alex Evilevitch$^{4,5}$,
Mandar M. Inamdar$^2$,
Prashant K. Purohit$^{2,6}$,\\
William M. Gelbart$^4$,
Charles M. Knobler$^4$,
and Rob Phillips$^{2,3}$ \\
  \small{$^1$ Department of Physics, $^2$ Division of Engineering and
    Applied Science,} \\
  \small{and $^3$ Kavli Nanoscience Institute,
    California Institute of Technology, Pasadena, CA;} \\
  \small{$^4$ Department of Chemistry and Biochemistry, 
    University of California, Los Angeles;} \\
  \small{$^5$ Department of Biochemistry, Center for Chemistry and Chemical
Engineering, Lund University, Lund, Sweden;} \\
  \small{$^6$ Department of Physics and Astronomy, University of Pennsylvania, Philadelphia.} \\
  \small{$^\dagger$ Address correspondence to: grayson@caltech.edu} \\
}
\maketitle

\begin{abstract}
A variety of viruses tightly pack their genetic material into protein
capsids that are barely large enough to enclose the genome.  In
particular, in bacteriophages, forces as high as 60~{\pN} are
encountered during packaging and ejection, produced by DNA bending
elasticity and self-interactions.  The high forces are believed to be
important for the ejection process, though the extent of their
involvement is not yet clear.  As a result, there is a need for
quantitative models and experiments that reveal the nature of the
forces relevant to DNA ejection.  Here we report measurements of the
ejection forces for two different mutants of bacteriophage
{\lamphage}, {\lamphageBlong} and {\lamphageC}, which differ in
genome length by ${\sim}30\%$.  As expected for a force-driven
ejection mechanism, the osmotic pressure at which DNA release is
completely inhibited varies with the genome length: we find 
inhibition pressures of 15~{\atm} and 25~{\atm}, respectively,
values that are in agreement with our theoretical calculations.
\end{abstract}

\section{Introduction}

Over the past thirty years, a series of experiments and theoretical
work have produced many insights about the importance of internal
forces in the
bacteriophage life cycle: Early measurements on {\lamphage} capsids
showed that they contained tightly packed DNA~\citep{earnshaw77}, and
subsequent experiments established that DNA packed at these densities
exerts a pressure of tens of atmospheres that is dependent on the
density and salt conditions~\citep{rau84}.  Any effect of the
{\lamphage} genome length on its life cycle (independent of any
particular genes) suggests that internal forces are important,
and there are several such effects known: there are upper and lower bounds
on the amount of DNA that can be packaged into a {\lamphage}
capsid~\citep{feiss77}; mutants of {\lamphage} with long genomes fail
to grow without magnesium ions~\citep{arber83}; and mutants with short
genomes fail to infect \textit{pel$^-$} cells~\citep{katsura83}.
While magnesium ions reduce the forces between DNA, stabilizing the
phage particles, DNA-condensing ions such as putrescine prevent DNA
ejection~\citep{katsura83}, and osmotic stress can stabilize
the genome within phages~\citep{serwer1983}.  The evidence seems to
indicate that
internal forces have an important role in the function of~{\lamphage}:
phages with low forces can be incapable of ejecting their DNA
forcefully enough to penetrate the cell, while phages with high forces
are unable to package their
genome or are unstable when fully packaged.
A variety of theoretical models of DNA
packaging in bacteriophage have been
proposed~\citep{riemer78, black89,serwer88,tzlil03,purohit03}, but
only recently have experiments begun to quantify the
forces required to tightly pack DNA into
capsids~\citep{smith01} and the forces driving DNA ejection
\citep{evilevitch03,evilevitch05}.

The aim of this paper is to study the effect of genome length on the
ejection force of {\lamphage} DNA, by comparing {\lamphageC}, a simple
mutant of the wild type with a 48.5 kbp genome, to {\lamphageBlong}
({\lamphageB}), which has a much shorter genome of 37.7
kbp~\citep{bellett71}.  The reason that measurements with different
genome lengths are especially interesting is that simple models of
the forces that arise during packaging depend in a precise way upon the
genome length.  To measure the force, a method reported
earlier~\citep{evilevitch03} was used: osmotic stress was applied to
the outside of the capsids during ejection, halting the ejection at
the point where the internal and external forces balance.  Using this
method, we show that phages with shorter genomes have lower forces,
phages with longer genomes eject their DNA with higher forces, and
that straightforward theoretical models are sufficient to predict
these effects.

\section{Materials and Methods}

\begin{figure}
\centering
\includegraphics[width=2in]{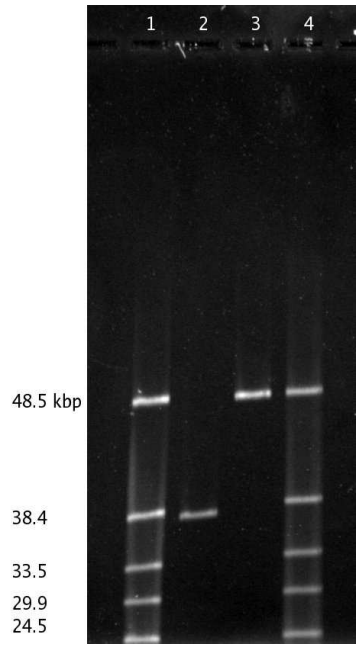}
\caption{
Comparison of {\lamphageB} (lane 2) and {\lamphageC} (lane 3)
  genomes by field-inversion gel electrophoresis.  The gel was run
  with 100/60~V switching for 19h in 0.5x TBE buffer, 1\% agarose.
  Lanes 1 and 4 are $\lambda$-mix ladders, with known lengths as
  shown.  Bands contain 0.5--1~ng of DNA, stained with SYBR Gold. The
  gel shows a single band in lane 2 at $37.9\pm 0.3$~kbp and a single
  band in lane 3 at $48.4\pm0.3$~kbp; both results are consistent with
  the expected values of 37.7~kbp and 48.5~kbp.
  }\label{compare-gel}
\end{figure}

\begin{figure}
\centering
\includegraphics[angle=-90,width=6in]{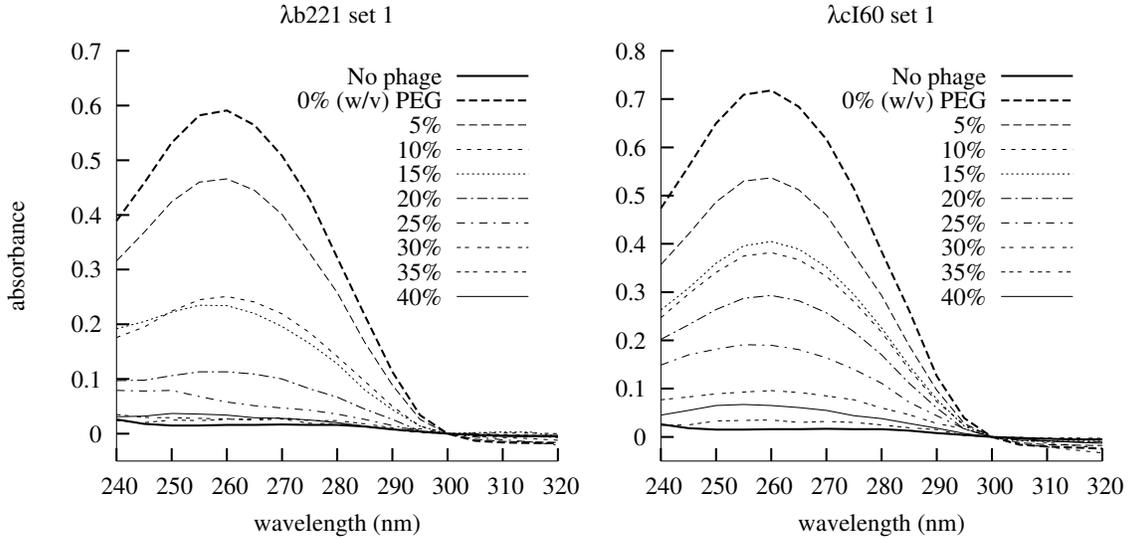}
\caption{
UV~absorbance of the supernatant containing fragments of the
  DNA ejected from {\lamphageB} (data set ``$\lambda$b221 set 1'', left) and
  {\lamphageC} (data set ``$\lambda$cI60 set 1'', right),
  aligned at 300~\nm.  Absorbance curves are shown for various
  concentrations of PEG, corresponding to different osmotic pressures,
  and for tests done in the absence of LamB.  A second data set taken
  for each phage is not shown.}\label{abs}
\end{figure}

Phages {\lamphageBlong} ({\lamphageB}) and {\lamphageC}
were extracted from single plaques, grown in 3~L cultures of {\ecoli}
c600 cells, and purified by PEG precipitation, differential
sedimentation, and equilibrium CsCl gradients, resulting in
${\sim}10^{13}$ infectious particles.  After purification, phages were
dialyzed twice against a 500-fold greater volume of TM buffer
(50~mM~Tris, 10~mM~$\rm MgSO_{4}$, pH~7.4).

To check the genome lengths of the phages used in this experiment, we
extracted the DNA with phenol and chloroform from approximately
$5\times 10^{9}$ phages of each type into 500~{\ul} of 0.5x TBE buffer.  A
quantity of 1~{\ul} ($10^7$ genomes, or 0.5~ng) was removed from each
extraction, mixed with loading dye, heated briefly to 65{\degC} to
separate cohesive ends, and pipetted into a 0.5x TBE, 1\% agarose
gel.  A 10~ng quantity of a standard ladder ($\lambda$-mix,
Fermentas) was included for size comparison.  We ran the gel using the
method of~\citep{birren90}, with a simple electrophoresis box (Owl
Separation Systems B1A) and a homebuilt voltage inverter, pulsed at
100~V forward, 60~V backward, for 19~h.  The gel was stained with SYBR
Gold (Molecular Probes) and photographed with an Alphaimager HP (Alpha
Innotech).  Dust was manually erased from the resulting image, which
is shown in
Fig.~\ref{compare-gel}.  Results were consistent with the expected
37.7~kbp genome for {\lamphageB} and 48.5~kbp genome for {\lamphageC}.

The {\lamphage} receptor LamB (maltoporin), required to trigger
ejection, was extracted from the membranes of {\ecoli} pop154 cells.
These cells express a \textit{lamB} gene from {\shigella} known to be
compatible with a variety of {\lamphage} strains, allowing ejection in
the absence of chloroform~\citep{roa1976,graff2002}.  LamB was affinity-purified in amylose
resin and dialyzed twice against TM buffer containing {1\%}
n-octyl-oligo-oxyethylene (oPOE; Alexis Biochemicals
\#500-002-L005.)

Our method for measuring ejection forces is substantially the same as
that described earlier~\citep{evilevitch03}, with minor refinements
that have improved precision.  We paid particular attention to the
difficulty of pipetting the viscous PEG solutions, trying to minimize
systematic and statistical errors that occur when the solution adheres
to the pipette tips.  A solution of 50\%~(w/w) polyethylene glycol
(PEG) 8000 (Fluka PEG Ultra) was prepared in TM buffer with 0.5\%
oPOE, and its density was measured at 1.09~g/ml (see
also~\citep{gonzalez-tello1994}).  This solution was used to prepare
solutions of PEG / TM 0.5--1\% oPOE at various specified \%(w/v)
values (see Fig.~\ref{abs}) on an analytical balance.  The mass
measurements allowed us to set the quantity of PEG in each sample
within 0.2~mg, which corresponds to an error in concentration of
approximately 0.1\%~(w/v).
Phage solution was added to a final concentration of
${\sim}10^{11}$/ml, and DNase~I was added at 10~\ug/ml.  The sample
tubes were turned slowly for several minutes to mix the viscous PEG
solutions.  Purified LamB was added with a wide-mouth tip and the
resulting {200~\ul} solution was mixed quickly by pipetting.
A wide-mouth tip is an inaccurate pipetting device, but it is
necessary for quickly mixing viscous solutions.  To minimize the
effect of inaccurate pipetting on the measurements, we used a
concentration of LamB that was sufficient for maximal
ejection, estimated at ${\sim}5$~\ug/ml
by UV absorbance.  After the addition of LamB, the samples were
incubated for 1~h at 37\degC, which was sufficient for the reaction to
reach its endpoint---complete digestion of the ejected genome fraction
by DNase~I.
Finally, the capsids were separated from the ejected DNA fragments by a
centrifugation for 20~h at $18{,}000\times g$.

After centrifugation, {120~\ul} of supernatant from each tube was
removed to a UV-transparent plastic cuvette (Ocean Optics UVettes) and
DNA concentrations were measured with a UV--visible spectrophotometer
(LKB Biochrom Ultrospec II).  The absorbance curves were aligned at
300~nm (immediately after the DNA absorbance peak) to compensate for
absorbance not due to DNA.  The resulting curves are shown in
Fig.~\ref{abs}.  The absorbance values at 260~nm, $A_{260}$, are
linearly related to the amount of DNA ejected from the phage
capsids.  In contrast with earlier experiments, there was no
measurable background DNA absorbance due to ruptured phage capsids:
samples at high PEG concentrations or without LamB had similar values
of $A_{260}$ to samples prepared without phages.  This is probably
because phages were used within one month of dialysis; in contrast,
samples of {\lamphageB} measured after five months of storage in TM
buffer at 4{\degC} had a background $A_{260} \approx 0.1$.

When there is no PEG present, ejection reaches 
completion (100\%), and when no phages are added there is no
ejection~(0\%).  Intermediate values were found with a linear
interpolation:
\begin{equation}
 \textrm{ejected fraction} =
  100\% \cdot ( A_\textrm{260,with PEG} - A_\textrm{260,no phages} ) /
              ( A_\textrm{260,no PEG} - A_\textrm{260,no phages} )\,.
\end{equation}
An alternative procedure is to use no LamB as a calibration for
0\%~\citep{evilevitch05}. 

The weight measurements set $\mPEG$, the mass of PEG, and $\mtot$, the
total mass, for each sample.  The \%~(w/w) weight-weight fraction was
computed as $w=\mPEG/\mtot$.  The osmotic pressure at each PEG
concentration was then determined with the empirical
formula~\citep{michel83}
\begin{equation}
\Pi(\atm) = -1.29 G^2 T + 140 G^2 + 4 G \,,
\label{pressure}\end{equation}
where T is the temperature (\degC) and $G \equiv w/(100-w)$.  Note
that the osmotic pressure is an increasing function of the PEG
concentration and a decreasing function of temperature.  For this
experiment, $T=37$.

Two complete sets of samples were prepared for each of {\lamphageC}
and {\lamphageB}.  Each sample has a statistical error due to weight
measurements, pipetting, and spectrophotometry.  To minimize
systematic effects on the ejected fraction, the two ``no phage'' and ``no
PEG'' tubes were averaged for each phage.  Statistical errors were
propagated to yield $x$ and $y$ error bars (see Fig.~\ref{results}.)

\section{Theoretical Model of DNA Packaging}

Our theoretical model is based on earlier work~\citep{riemer78,
tzlil03, purohit03, purohit05}, which describes the packaging energy
as a function of the length of DNA in the capsid.  We model the
{\lamphage} capsid as a sphere  and its genome as a long semiflexible
rod.  We
assume that the rod is wound into a cylindrically symmetric
spool~\citep{cerritelli97} with local hexagonal packing.  The total
energy of the packaged DNA can then be approximated by a sum of
inter-axial repulsion energy and the bending energy of the rod:
\begin{equation}\label{packingenergy}
E = E_{\rm interaction} + E_{\rm bend} = \sqrt{3} F_0L(c^2 +
cd_s)\exp(-d_s/c) + \pi \kB T\xi
\int_{R_{\rm in}}^{R_{\rm out}} {N(r)\over r} dr \,,
\end{equation}
where $F_0$ and $c$ are empirically determined constants describing
the interaction between neighboring DNA double-helices, $\xi$ is the
persistence length of DNA, $L$ is the length of the DNA within the
capsid, $d_s$ is the inter-axial spacing, $\Rout$ and $\Rin$ are the
radius of the capsid and the inner radius of the DNA spool,
respectively, and $N(r)$ is the number of loops of DNA at a distance
$r$ from the spool axis.  For the persistence length $\xi$ we use
50~nm, though its value in Mg$^{2+}$ buffer may be~$\sim$10\%
smaller~\citep{hagerman1988}.  The spacing between sequential bases
of DNA varies, depending on the base types, from 0.33 to
0.34~$\nm$~\citep{olson1998}.  To compute $L$ we disregard this
variation and use 0.34~$\nm$ times the number of base pairs within the
capsid.  The inter-axial forces in buffers containing
$\rm Mg^{2+}$ have been measured~\citep{rau84}.  Since the values
measured for 5~mM and 25~mM $\rm Mg^{2+}$ were not significantly
different, we assume that the forces at 10~mM (used in our experiments)
will be identical.  A least-squares fit to the 5~mM and 25~mM data
in~\citep{rau84} gives $F_0 = \hbox{12,000}$~$\pN/\nm^{2}$ and $c =
0.30~\nm$.
The radius of the phage capsid $\Rout$ is around $29$
nm~\citep{earnshaw77}.  Once we know $d_s$, $\Rin$, and $N(r)$,
we can use Eq.~\ref{packingenergy} to calculate the
internal force
on the phage genome as a function of genome length inside the capsid,
providing an interpretation of the experimental results.

We calculate the remaining variables as a function of $L$ according to
the following recipe, which
involves only simple geometrical considerations and elementary
calculus. The number of loops $N(r)$ in Eq.~\ref{packingenergy}
is given by $z(r)/d_s$, where $z(r) = \sqrt{\Rout^2-r^2}$ is the height of the capsid at
distance $r$ from the central axis of the DNA spool. The actual volume
$V(R_{\rm in}, R_{\rm out})$ occupied by the DNA spool can be related
to the genome length $L$ and the inter-axial spacing $d_s$ to get an
expression for $R_{\rm in}$ in terms of $d_s$, $R_{\rm out}$ and
$L$~\citep{purohit03}. This expression for $R_{\rm in}$ is substituted
into Eq.~\ref{packingenergy}, which can then be minimized with
respect to $d_s$ to give the equilibrium inter-axial spacing as a
function of the genome length $L$ inside the capsid.  From $d_s$, $\Rin$,
and~$N(r)$,
Eq.~\ref{packingenergy} now gives us the total packing energy as
a function of genome length inside the capsid. The internal force
$F(L)$ acting on the genome is obtained by taking the derivative of
Eq.~\ref{packingenergy} with respect to $L$~\citep{purohit03}.

The preceding construct is a parameter-free model that predicts the
ejection force from a {\lamphage} capsid.  Experimental uncertainties
in the quantities quoted above should lead to errors of $10$--$50\%$ in
the magnitude of the force predicted, but the shape and relative
positions of the curves for different genome lengths should not be 
strongly affected by these errors (these tests of the parameters are
not shown).

DNA ejection in our experiment or \textit{in vivo}
is halted at the point where the internal force balances the
osmotic force.  We have described above how to obtain the internal
\emph{force} (and hence, the external osmotic force because of the
equilibrium) acting on the genome. But the experimental variable is an
osmotic \emph{pressure}~\citep{tzlil03}. Thus we need to translate this
force into a pressure. The force $F(L)$ is given approximately by
$\Pi\cdot \pi \RDNA^2$, where $\Pi$ is the osmotic
pressure and $\RDNA$ is the effective radius of the DNA.  We take
$\RDNA$ as 1.0~nm (bare DNA) plus 0.2~nm, half the PEG monomer length
found experimentally~\citep{abbot92,marsh04}.
This is an exact formula for the osmotic force on a very large area.
However, at the PEG concentrations used in our experiment, the
diameter of DNA is comparable to the correlation length
(mesh size) of PEG, $\sim$1--3~{\nm}~\cite[pp.78--80]{degennes79}, so
the formula for $F(L)$ is only approximately valid.  
A scaling expression for the correction could be
used~\citep{devries01,castelnovo03,evilevitch04a},
but since this result is good only
up to a multiplicative constant, its importance is
unclear. Hence, to maintain clarity in our analysis we do not use the
correction.  On the other hand, including this effect could result in
a better fit between theory and experiment.

\section{Results and Discussion}
\begin{figure} 
\centering
\includegraphics[angle=-90,width=6in]{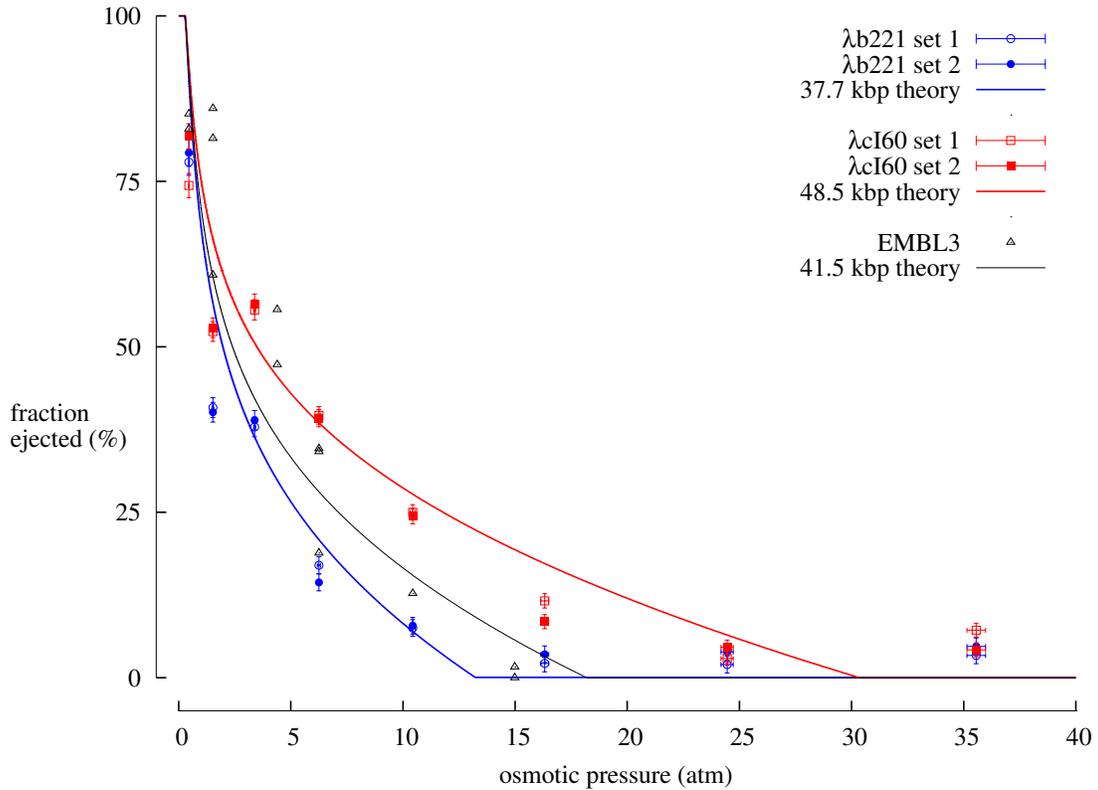}
\caption{
Measured ejection percentages for four sets of data on {\lamphageC}
and~{\lamphageB} (points) compared to theoretical
predictions (curves) for 37.7~kbp and 48.5~kbp genomes.  Error
bars showing one standard deviation were computed from weight
measurement, pipetting, and spectrophotometer errors.  At all
pressures below 15~atm, {\lamphageC} ejects more DNA than {\lamphageB}.
There is an interesting ``dip'' in the experimental data
for both phages, between 1.5 and 3.5~\atm, but otherwise the precise
trends predicted by theory are seen in the data.  The theoretical
pressures are calculated with the
approximate relation $\Pi\, (\atm) = F / \pi \RDNA^2$, where $\RDNA$
is the effective radius of DNA, assumed to be 1.2~nm.
Also plotted: data reported earlier on EMBL3~\citep{evilevitch03},
with corrections made for a systematic error in the earlier analysis.
There is a larger scattering error in the EMBL3 data than in the
{\lamphageC} and~{\lamphageB} data; taking this
into consideration, the data sets for all three phages are completely
consistent with the expectation that phages with longer genomes eject
more DNA at all pressures.
}
\label{results}
\end{figure}

Fig.~\ref{results} shows our experimental results.
Partial ejection in both {\lamphageB} (37.7~kbp) and {\lamphageC}
(48.5~kbp) corresponds closely to that measured earlier in EMBL3
(41.5~kbp): the ejected fraction decreases quickly to 50\% at several
atmospheres of osmotic pressure, then descends more slowly to 0\%.
(As shown by~\cite{evilevitch05}, this ``ejected fraction''
corresponds to all phages ejecting the same percentage of their DNA,
rather than a percentage of the phages ejecting all of their DNA.)
Most importantly, at 
every pressure where a measurable amount of DNA is ejected,
{\lamphageB} ejects less DNA than {\lamphageC}.  Conversely, at any
point in the ejection process, the osmotic force on the DNA is always
lower in {\lamphageB} than in {\lamphageC}.  For example, the
highest forces occur at the point where osmotic pressure
completely inhibits the ejection of DNA.  This inhibition pressure is
in the range 10--15~{\atm} for {\lamphageB} and 20--25~{\atm} for
{\lamphageC} --- a $\sim$100\% increase in pressure results from a 30\%
increase in genome length.

An interesting dip (or bump) in the data is apparent between 1.5 and
3.4~\atm. We observed this dip in all data sets collected for this
experiment, but its cause is unknown.

Fig.~\ref{results} also shows the predictions of our parameter-free
theory.  Except for the dip, the theory predicts the data quite well.
Both the absolute magnitude and the general shape of the curves are
predicted correctly.  This is a remarkable result
because no fitting was done to match the theory to
the data.  A straightforward inverse-spool model of the DNA
arrangement, taking into account only the measured bending
elasticity and interaxial forces, correctly predicts the dependence of
ejection force on genome length.

Our measurements can be compared with previous results on
EMBL3~\citep{evilevitch03}, for which the genome length is
41.5~kbp. To do so, however, requires that the weight fractions be
corrected in calculating the osmotic pressures, which we have done
here for the rest of the data. In this earlier work, the osmotic
pressures of the solutions were calculated from the \%(w/v)
concentrations, without taking into account the density of the PEG
solution.  Here, we convert it to \%(w/w) taking into account the
density of the PEG solution. The the density differences can be
corrected using the relation between weight fraction and density of
PEG solutions~\citep{gonzalez-tello1994}.  After this correction, we
find that the maximum pressure, for example, is
15~atm instead of 19.6~atm. The correction has a smaller effect at
lower PEG concentrations; the difference at 10\% (w/v) is only 0.2
atm.  Fig.~\ref{results} shows a comparison between EMBL3 and the
mutants used in the present study. Within the experimental
uncertainty, the ejection curve for the intermediate length genome
lies between those of the 48.5~kbp and 37.7~kbp phages reported here.

\begin{figure} 
\centering
\includegraphics[angle=-90,width=6in]{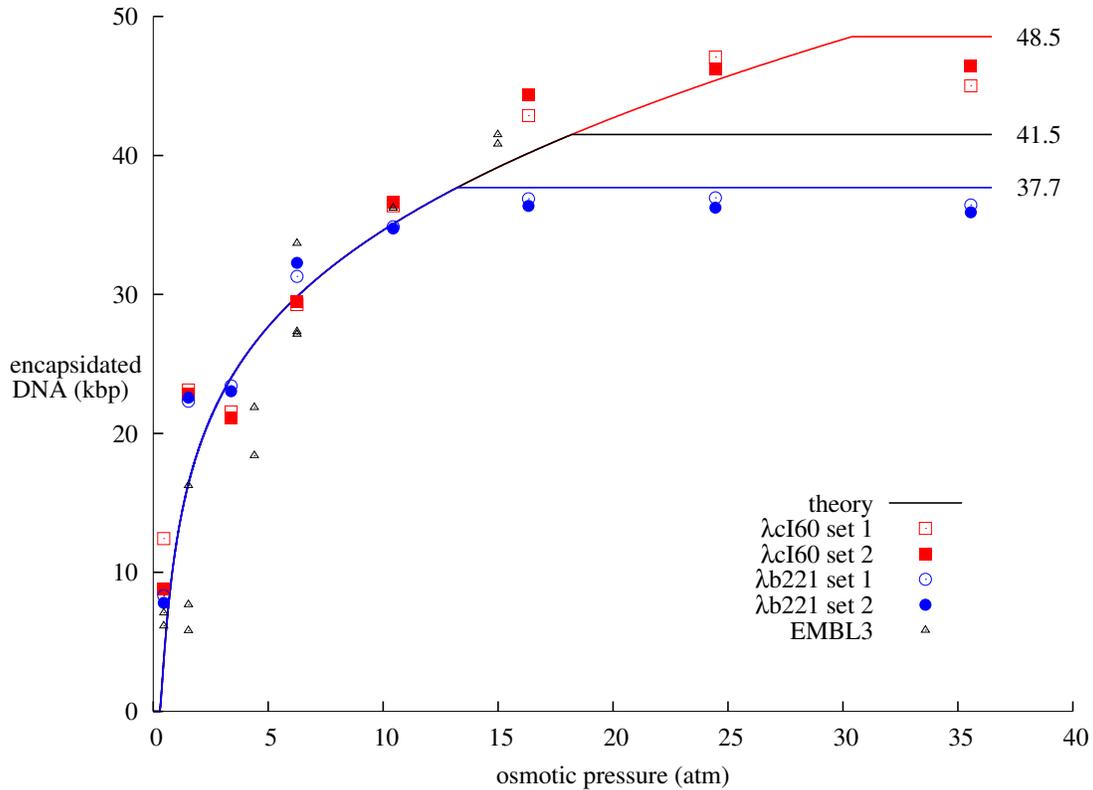}
\caption{
Measured ejection data, plotted as the amount of DNA remaining in the
capsid after ejection.  The remaining DNA is calculated as $N\cdot(1 - p)$,
where $N$ is the number of base pairs in the genome and $p$ is the
ejected fraction plotted in Fig.~\ref{results}.  Dotted
lines at 48.5, 41.5, and 37.7~kbp correspond to the full genomes of
{\lamphageC}, {\lamphageB}, and EMBL3, respectively.  As
expected, the data points fall on a single curve below 10~atm, then
diverge as the maximum pressure is reached for each phage.
}
\label{results_inverted}
\end{figure}

An alternative view of the ejection data is shown in
Fig.~\ref{results_inverted}.  From the ejected fraction~$p$ plotted
in Fig.~\ref{results}, we can compute
the amount of DNA remaining in the capsid as $N\cdot(1 - p)$, where $N$ is
the number of base pairs in the genome.  Since ejected DNA is digested
by DNase~I, the amount remaining in the capsid should depend only on
the external osmotic pressure, and it should be independent of the
genome size of the phage.  Indeed, Fig.~\ref{results_inverted} shows
that all of the data below 15~atm falls roughly on a single curve.
Above 25~atm, the pressure is sufficient to hold the entire genome
within the capsid, so the amount of encapsidated DNA is just equal to
the genome length of the phage.

For {\lamphageC}, which is almost identical to the wild-type
{\lamphage}, our model and measurements indicate that DNA is ejected
with a force of around 10~pN that drops steadily as the genome enters
the host cell.  What advantage does {\lamphage} gain from having such
a high ejection force?  For the phage DNA to enter the {\ecoli} cell,
it must overcome an internal osmotic pressure of
$3~\atm$~\citep{neidhardt96}.  Fig.~\ref{results} shows that
{\lamphageC} can eject roughly 60\% of its genome, 30~kbp, against this
osmotic pressure.  However, the rest of the genome must be actively
transported into the cell.  Reversible diffusion, in particular, is
incapable of transporting the rest of the genome into cell, since the
energy barrier that must be overcome is several times $10^3 \kB
T$~\citep{smith01,purohit05}.  A two-step process is implied: first a
quick pressure-driven injection of half of the DNA, then a slower
protein-driven importation of the remainder (as, for example, by RNA
polymerase 
in phage T7).  This two-step process has been observed \textit{in
vivo} for the phages T5~\citep{letellier04} and
\phiphage~\citep{gonzalez-huici04}, though similar experiments 
have not been done on phage \lamphage.  For {\lamphageB},
Fig.~\ref{results} shows that only 40\% of the genome, 15~kbp, is
injected in the first step.  Since phages with shorter genomes than
{\lamphageC} are not infectious, we speculate that 15~kbp is close to
the minimum amount of DNA required to initiate the second
(non-pressure-driven) stage of importation.

In the case of bacteriophage T7, however, the entire genome is
actively transported at a slow, constant speed~\citep{molineux01,kemp04},
even though calculations indicate its internal pressure to be
similar to that in {\lamphage}~\citep{purohit05}.  Apparently,
something unique to T7 slows down the process (perhaps the observed
expulsion of internal proteins from the T7 capsid?), whereas most dsDNA
bacteriophages experience a first stage of fast, pressure-driven
ejection.  Since ejection from {\lamphage} is much
faster~\citep{novick1988,kemp04} than from T7, it seems that {\lamphage} is
in this most common category.  (As another example, recent experiments
by~\cite{frutos2005} and~\cite{mangenot2005} show that DNA ejection from
bacteriophage T5 proceeds very rapidly between discrete stopping
points along the genome.  Furthermore, it has been shown in
unpublished measurements that T5 ejection can be inhibited by external
osmotic pressure.) \textit{In vivo} ejection
experiments with {\lamphage} and \textit{in vitro} experiments with T7
need to be done to explore the differences between these two phages.

\section{Acknowledgments}
We thank Alexandra Graff and Emir Berkane for providing protocols for
the purification of LamB and the pop154 {\ecoli} strain.  Michael
Feiss kindly sent us samples of the {\lamphageB} and {\lamphageC}
phages used here.  We are indebted to Douglas Rees, Scott Fraser,
Stephen Quake, and Grant Jensen for laboratory space and equipment; and
to Ian Molineux, Jonathan Widom, and others for very helpful
conversations.
This work was supported by a grant from the Keck Foundation (to RP), an NIH
Director's Pioneer Award (to RP), NSF grant CMS-0301657 (to RP), and
NSF grant CHE-0400363 (to CMK and WMG).  PG was supported by an NSF
graduate research 
fellowship.  AE has received financial support from The Swedish
Foundation for International Cooperation in Research and Higher
Education (STINT) and the Swedish Research Council (VR).

\bibliographystyle{mmb}
\bibliography{paper}

\end{document}